# Increased Curie temperature and enhanced perpendicular magneto anisotropy of $Cr_2Ge_2Te_6$/NiO heterostructure

H. Idzuchi[1,2,#,*], A. E. Llacsahuanga Allcca[2,#], X. C. Pan[1], K. Tanigaki[1,4] and Y. P. Chen[2,1,3,5]

[1] *WPI Advanced Institute for Materials Research (AIMR), Tohoku University*

*Sendai 980-8577, Japan*

[2] *Purdue Quantum Science and Engineering Institute and Department of Physics and Astronomy,*

*Purdue University, West Lafayette, Indiana 47907, USA*

[3] *Center for Science and Innovation in Spintronics, Tohoku University*

*Sendai 980-8577, Japan*

[4] *Department of Physics, Graduate School of Science, Tohoku University*

*Sendai, 980-8578*

[5] *School of Electrical and Computer Engineering and Birck Nanotechnology Center,*

*Purdue University, West Lafayette, Indiana 47907, USA*

[*]idzuchi@tohoku.ac.jp

[#]) H. Idzuchi and A. E. Llacsahuanga Allcca contributed equally to this work.

**Abstract**

Magnetism in two-dimensional van der Waals materials has received significant attention recently. The Curie temperature reported for those materials, however, has been so far remained relatively low. Here, we measure magneto-optical Kerr effects (MOKE) under perpendicular magnetic field for van der Waals ferromagnet $Cr_2Ge_2Te_6$ as well as its heterostructure with antiferromagnetic insulator NiO. We observe a notable increase in both Curie temperature and magnetic perpendicular anisotropy in $Cr_2Ge_2Te_6$/NiO heterostructures compared to those in $Cr_2Ge_2Te_6$. Measurements on the same exfoliated $Cr_2Ge_2Te_6$ flake (on a $SiO_2$/Si substrate) before and after depositing NiO show that the hysteresis loop can change into a square shape with larger coercive field for $Cr_2Ge_2Te_6$/NiO. The maximum Curie temperature ($T_C$) observed for $Cr_2Ge_2Te_6$/NiO reaches ~120 K, is nearly twice the maximum $T_C$ ~ 60 K reported for $Cr_2Ge_2Te_6$ alone. Both enhanced perpendicular anisotropy and increased Curie temperature are observed for $Cr_2Ge_2Te_6$ flakes with a variety of thicknesses ranging from ~5 nm to ~200 nm. The results indicate that magnetic properties of two-dimensional van der Waals magnets can be engineered and controlled by using the heterostructure interface with other materials.

Recently, magnetism in layered van der Waals (vdW) materials has attracted large attention because of their unique magnetic properties stemming from their two-dimensional (2D) nature [1, 2]. Such layered vdW materials provide an opportunity to fabricate hetero-structures free from constraints in conventional film growth, and promise a unique route to explore new functionality of these materials based on electric field and crystalline symmetry [3,4]. One of the major issues in vdW ferromagnets is that the ferromagnetic transition temperature (Curie temperature) is relatively low. $CrI_3$ and $Cr_2Ge_2Te_6$ were reported as atomically-thin-form ferromagnets in 2017, where the Curie temperature ranges from 30 K (bilayer $Cr_2Ge_2Te_6$) to 45 K (monolayer $CrI_3$), being intriguingly low compared to the bulk values of 61 K (bulk $CrI_3$) and 66 K (bulk $Cr_2Ge_2Te_6$) [1,2]. To enhance the Curie temperature, a variety of approaches using interface and gap engineering have been proposed including dielectric effect, spin-orbit coupling proximity, charge transfer and interface hybridization [5]. However, only a few approaches have been implemented and reported so far, such as electric gating [4,6]. Therefore, it is important to search for other effective approaches to enhance the Curie temperature. Here, we study magnetic properties in heterostructures between antiferromagnet NiO and vdW ferromagnet $Cr_2Ge_2Te_6$. We will report magneto optical Kerr effect (MOKE) and detect hysteresis arising from ferromagnetism.

$Cr_2Ge_2Te_6$ was earlier reported by Carteaux et al [7] and recently there is significantly increasing interest because this is a layered two-dimensional magnet with mechanical cleavability down to atomically thin layers. The $Cr_2Ge_2Te_6$ single crystals used in this work were grown via a self-flux technique. First, 100 mg of Cr powder, 200 mg of Ge powder and 2 g Te were sealed in a quartz tube. The mixture was heated to 1050 °C and held for 30 hours, then cooled down to 475 °C in 10 days, and finally the Ge-Te flux was removed using a centrifuge at this temperature. The magnetic properties of the bulk crystals were characterized by magnetometry using MPMS (magnetic properties measurement system), and the Curie temperature was found to be $T_C \approx 66$ K from the minimum of $dM/dT$ curves (measured with the magnetic field of 50 mT in the $c$-axis).

The observed magnetic properties were consistent with the previous reports [1,4,7-13]. $Cr_2Ge_2Te_6$ crystals were mechanically cleaved onto a silicon substrate in ambient conditions (with the $SiO_2$ thickness of 285 nm). The thickness of the flakes was characterized by an atomic force microscope. Atomically thin flakes were visible with thickness-dependent color contrast due to the interference effect as shown in Fig. 1(a). Sputtering was performed using a NiO target with a base pressure of ca. $1\times10^{-5}$ Pa, Ar pressure of 0.2 Pa, and 200 Watt RF power. Various deposited NiO films with different thicknesses of ranging from 20 nm to 100 nm are explored in this work. After the sputtering of NiO layer, the color of both silicon substrate (with NiO) and the $Cr_2Ge_2Te_6$ flakes changed as shown in Fig. 1(b) because the interference condition was modified.

Polar magneto-optical Kerr Effect (MOKE) measurements were performed in an Oxford MicrostatMO system in the Faraday configuration. A temperature stabilized laser diode (635 nm in wavelength) was used to deliver a linearly polarized laser beam that was focused onto the sample at normal incidence using a 0.6 NA 100X long working distance objective. The estimated power delivered to the sample is less than 3 µW. The laser beam was intensity modulated using a chopper and the reflected beam was sent through a Wollaston prism oriented at 45 degrees with respect to the initial polarization to split the beam into two. The split beams were collected by two photodiodes arranged in a differential mode whose output was sent to an SR570 current pre-amplifier (Stanford research systems). The output of the amplifier was then measured using a lock-in referenced to the chopper frequency. In order to complete the measurements, each photodiode was covered in turn to obtain the total intensity signal. This is used to normalize the previously measured signals that will yield a quantity proportional to the Kerr effect signal (rotation of the polarization angle). Calibration was performed to obtain the actual conversion factor between radians and normalized signal by tilting the polarizer by a few fractions of a degree and recording the normalized signal.

Figure 1c shows MOKE curves (Kerr rotation angle versus magnetic field) for a sample before NiO deposition. The magnetic field was applied perpendicular to the substrate because the easy axis of $Cr_2Ge_2Te_6$ was reported in that direction. For the flakes with the thicknesses more than 5 nm, we observed clear hysteresis. For the flakes with a thickness of 5 nm or less, we did not observe clear hysteresis, which can be attributed to the weaker two-dimensional magnetism [1], and/or to the possibly not-pristine surface or film degradation (e.g. oxidization). After 20-nm-thick NiO layer was deposited, we measured the same flakes for comparison. These flakes, for the ones that showed a clear hysteresis before the deposition, showed an increase of coercive field and a change of the hysteresis into a rectangular shape as shown in Fig.1(d). This clearly indicates the enhanced perpendicular anisotropy induced by depositing the NiO layer.

Importantly, $Cr_2Ge_2Te_6$/NiO not only enhances perpendicular anisotropy but also increases the Curie temperature. Figure 2a shows temperature dependence of the hysteresis curves in MOKE signal for the sample shown in Fig. 1 ($Cr_2Ge_2Te_6$ thickness 7 nm, position 2). We characterized the Curie temperature as the midpoint between the temperature in which the coercive field is no longer seen and the one in which we still see a coercive field. For the position 2, we observed that the Curie temperature increased by 30 K to ~85 K after NiO deposition. Similarly, we observed an increased Curie temperature to 85 K at the position 1 ($Cr_2Ge_2Te_6$ thickness 8 nm), to 70 K at the position 3 ($Cr_2Ge_2Te_6$ thickness 6 nm), and to 70 K at the position 4 ($Cr_2Ge_2Te_6$ thickness 5.5 nm). For comparison, we measured another sample with 50 nm thickness of NiO, while the thickness of the flake was similar to that of position 2. We observed an even larger increase of the Curie temperature up to 115 K as shown in Fig.2(b), which is about twice of the Curie temperature on the $Cr_2Ge_2Te_6$ flakes without NiO. In this sample, we also observed further enhanced perpendicular anisotropy as the coercive field was even higher than that of a sample with NiO thickness of 20 nm.

Figure 2c shows hysteresis of a 202-nm-thick $Cr_2Ge_2Te_6$ flake with NiO of 20 nm. The hysteresis loop at T = 7 K shows small perpendicular anisotropy (coercive field of nearly 0 Oe, similar to that measured in bulk $Cr_2Ge_2Te_6$, see the supplementary material). At elevated temperatures (T = 55 K and above, where ferromagnetism disappears in bulk $Cr_2Ge_2Te_6$ with typical $T_C \sim$ 60 K), the hysteresis becomes more rectangular in shape and indicates the presence of some NiO-induced enhancement of perpendicular anisotropy even for such thick flakes. We also raise the possibility in heterostructures of relatively thick $Cr_2Ge_2Te_6$ that we could be accessing to the (different) magnetism of $Cr_2Ge_2Te_6$/NiO interface (observed in the data above 55 K in Fig. 2c) separately from that of bulk $Cr_2Ge_2Te_6$ itself (observed in the data at 7 K). For the case of thin flakes, as exemplified in Fig. 2a and 2b, one likely cannot separate the signal from the $Cr_2Ge_2Te_6$/NiO interface and the bulk $Cr_2Ge_2Te_6$ itself, but rather should consider it as the whole $Cr_2Ge_2Te_6$ affected by the interface with NiO.

Figure 3 summarizes the Curie temperatures for $Cr_2Ge_2Te_6$ flakes with various thicknesses, both for those without and with NiO of 20 nm, 35 nm, 50 nm, and 100 nm in thickness together with the data in literature for $Cr_2Ge_2Te_6$ without NiO [1, 4]. The data clearly shows that $Cr_2Ge_2Te_6$/NiO can enhance the Curie temperature. For very thin $Cr_2Ge_2Te_6$ flakes (thickness less than 5 nm) we were not able to see a clear enhancement, possibly being attributed to the defective layers (we note that the monolayer is ~0.7 nm according to crystallographic study [7]). We avoided unnecessarily exposing the samples to the air (the exposure time before NiO deposition is typically within five minutes). For thin $Cr_2Ge_2Te_6$ flakes (thickness more than 5 nm and less than 30 nm), the increase of the Curie temperature is strong for the 50-nm-thick NiO. We note the different optical constant and the interference effect on Si/$SiO_2$/$Cr_2Ge_2Te_6$/NiO with different thickness of NiO layers, as the color of Si/$SiO_2$/$Cr_2Ge_2Te_6$/NiO is different (in fact the substrate color of Si/$SiO_2$/NiO is also clearly different upon the thicknesses of NiO layers, for example between 50-nm-thick NiO and 20-nm-thick NiO). In addition, the opposite sign of the MOKE hysteresis curves between Fig.1 (c) and (d) is supposedly due to the different optical constants [14]. While such difference in optical constant and interference (giving also different effective optical penetration depths) can affect the MOKE signal (sign and amplitude) due to the different NiO thickness, it is more difficult to explain the difference in magnetic properties (Tc and anisotropy) of the materials themselves. One of such possible factors would be the difference in crystalline quality and interfaces for different thickness of NiO layers. As a controlled experiment to probe this question, we characterized the crystallinity of NiO with different thickness for Si/SiO$_2$/Pt/NiO of 50-nm-thick NiO and 20-nm-thick NiO by X-ray diffraction (SmartLab, RIGAKU Corporation). The (111) and (002) peaks are clearly visible (see the supplementary material for the detail). The full width half maximum (FWHM) of (111) peak is 0.60 deg and 0.70 deg for the 50-nm and the 20-nm thick NiO, respectively. These results were concluded to be the 10-15 nm grain size depending on the details of analysis. The grain size between the two NiO's of different thickness varies only by ~15%. It is not clear presently if the difference in the Curie temperature of Cr$_2$Ge$_2$Te$_6$/NiO can be related to this relatively moderate difference in crystallinity. Another possible mechanism might be related to the strain, as we noticed that wrinkles appeared in Cr$_2$Ge$_2$Te$_6$ after the NiO deposition. This mechanism could be consistent with the middle panel of Fig. 3, where the thick Cr$_2$Ge$_2$Te$_6$ flakes (thickness of more than 30 nm) shows little difference of Curie temperature between NiO's of different thicknesses. For thick Cr$_2$Ge$_2$Te$_6$ flakes we still see the increase of Curie temperature as we discussed in the previous paragraph. The Curie temperature in thick flakes with NiO is higher than that of the bulk Curie temperature reported in various literatures [1, 4, 7-13] as shown in the right panel of Figure 3. This higher Tc in the thick flakes could mostly be contributed by the (enhanced) magnetism at the Cr$_2$Ge$_2$Te$_6$/NiO interface as we discussed earlier (Fig. 2c). While the precise mechanism of the increase in Curie temperature as well as the enhancement in perpendicular anisotropy for Cr$_2$Ge$_2$Te$_6$/NiO interface magnetism cannot be clear yet at this stage, so far there are a few reports on the correlations between the perpendicular anisotropy and the Curie temperature in other systems. Recently it has been reported that a two-dimensional magnet with additional perpendicular anisotropy shows an increase of the Curie temperature [15, 16].

Bonding between 3$d$ electrons in transition metal and 2$p$ electrons in oxygen is known to induce high perpendicular anisotropy for MgO/Fe, MgO/Co both from experimental and theoretical studies [17,18], which may provide a microscopic mechanism for the enhanced perpendicular anisotropy and increase of the Curie temperature observed in the present studies. We also note that we did not observe any exchange bias (for example reported earlier for NiO/Ni$_{81}$Fe$_{19}$ interface [19]) in our hysteresis loop. It is noted that magnetic field was not applied when we made the interface during the NiO deposition.

In summary, we studied magnetic properties of a two-dimensional van der Waals magnet, Cr$_2$Ge$_2$Te$_6$ flakes covered by NiO thin films by magneto optical Kerr effects. We characterized the hysteresis loops of Cr$_2$Ge$_2$Te$_6$ flakes before and after NiO deposition. Cr$_2$Ge$_2$Te$_6$/NiO showed a strong increase in Curie temperature and a clear enhancement in perpendicular anisotropy as evidenced by the increase of coercive field and the change of hysteresis into a rectangular shape. We observed the Curie temperature as high as 115 K for Cr$_2$Ge$_2$Te$_6$/NiO with NiO thickness of 50 nm, more than twice the one for Cr$_2$Ge$_2$Te$_6$ without NiO. The Curie temperature increase for Cr$_2$Ge$_2$Te$_6$/NiO is observed for Cr$_2$Ge$_2$Te$_6$ with thicknesses ranging from 5 nm to 200 nm. Even Cr$_2$Ge$_2$Te$_6$/NiO with 200 nm thick Cr$_2$Ge$_2$Te$_6$ flakes showed the higher Curie temperature than that of bulk Cr$_2$Ge$_2$Te$_6$. These results indicate magnetic properties of two dimensional van der Waals materials can be controlled by employing hetero-structure and interface with other materials.

**Supplementary material**

See supplementary material for the hysteresis curve of bulk Cr$_2$Ge$_2$Te$_6$ and XRD data of Si/SiO$_2$/Pt/NiO.

**Acknowledgments**

We acknowledge K. Saito for the growth of NiO thin film and XRD measurement, and P.

Upadhyaya, D.Xiao, A. Rustagi, T. Nakanishi and A. Lu for fruitful discussions. This work was supported in part by a Grant-in-Aid for Scientific Research from the Ministry of Education, Culture, Sports, Science and Technology (MEXT), JSPS KAKENHI (Grant Number 18H03858, 18H04473, 18H05840, 18H04471, 17-18H05326, 18H04304, 18H03883, 18F18328), Sumitomo Foundation (Grant Number 180953), WPI‐AIMR's fusion research program under World Premier International Research Center Initiative (WPI), MEXT, Japan, Center for Science and Innovation in Spintronics and Inter-University Cooperative Research Program of the Institute for Materials Research, Tohoku University (Proposal Number 19G0210, Cooperative Research and Development Center for Advanced Materials), Purdue University and National Science Foundation (Grant DMR1838513). Xing-Chen Pan acknowledges support from an International Research Fellowship of Japan Society for the Promotion of Science (Postdoctoral Fellowships for Research in Japan (Standard)).

**Figures**

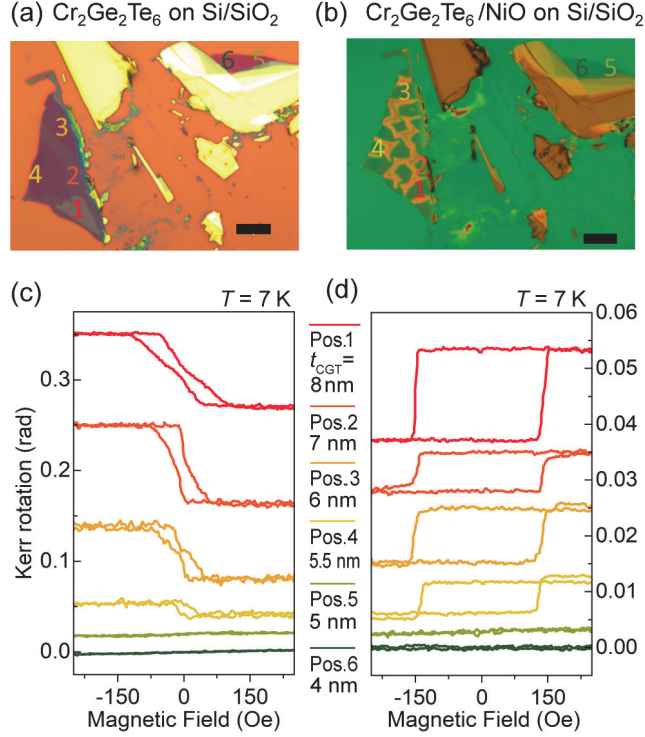

**FIG. 1.** (a), (b) Optical microscope image of Cr$_2$Ge$_2$Te$_6$ (CGT) flakes on Si/SiO$_2$ substrate, before (a) and after (b) deposition of NiO. After deposition of NiO with thickness of 20 nm, the optical contrast changes because the interference condition is modified. Scale bar is 10 μm. (c) Measured magneto optical Kerr effect (MOKE) curves at the temperature of 7 K, with the magnetic field perpendicular to the substrate. The curves from the different positions (labeled 1-6) are shifted vertically. The positions are marked in the images (a) and (b). The CGT-flake thickness at each position was measured by atomic force microscopy. (d) MOKE curves measured at the same positions after NiO deposition. Perpendicular anisotropy is strongly enhanced and square shaped hysteresis curves are observed. The opposite sign of the MOKE hysteresis curves between (c) and (d) is supposedly due to the different optical constants of the sample without (c) and with (d) the NiO overlayer [14].

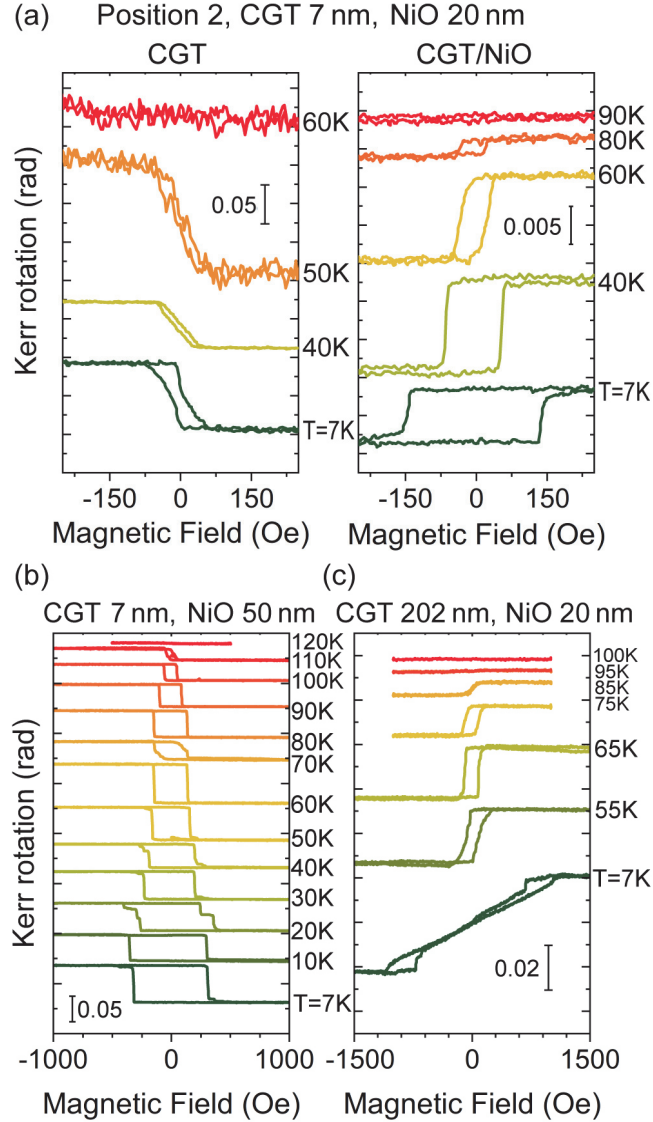

**FIG. 2.** Temperature dependence of MOKE curves: (a) Comparison for the same flake/position (position 2 in Fig. 1ab) before and after NiO deposition. The thickness of Cr$_2$Ge$_2$Te$_6$ (CGT) is 7 nm and the thickness of NiO is 20 nm. After NiO deposition, the Curie temperature is increased from ~55 K to ~85 K. (b) The signal measured on a different Cr$_2$Ge$_2$Te$_6$ flake with similar thickness but thicker NiO (thickness 50 nm), showing even stronger increase of Curie temperature to ~115 K. (c) The MOKE signal measured on a relatively thick (202-nm-thick) Cr$_2$Ge$_2$Te$_6$ flake with NiO with the same thickness of 20 nm as shown in (a), also showing an enhancement of Curie temperature to ~90 K.

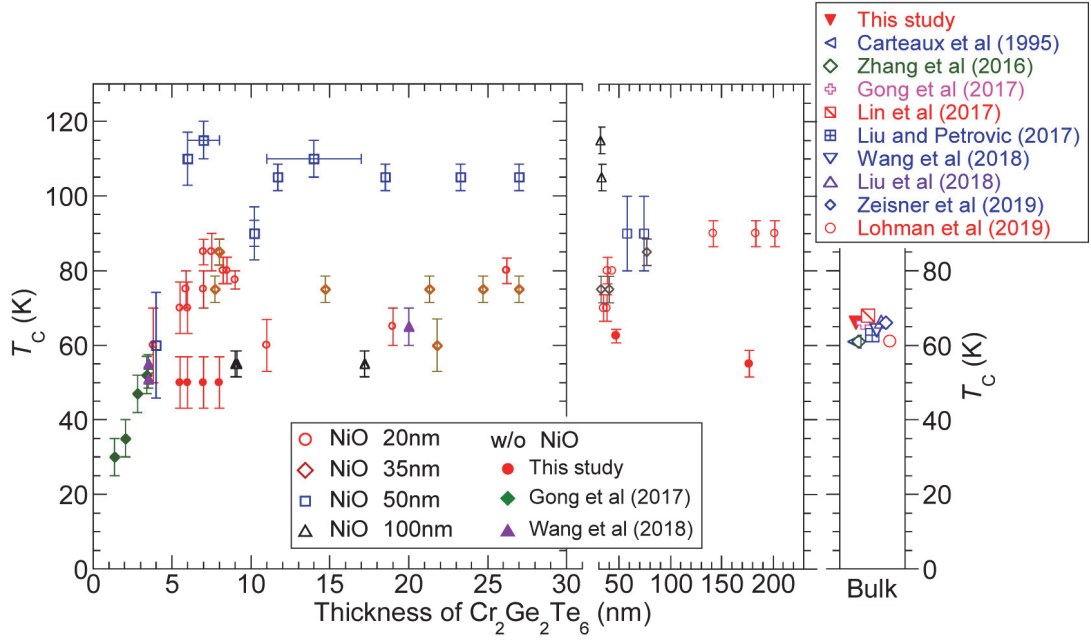

**FIG. 3.** (color online) Curie temperatures measured on $Cr_2Ge_2Te_6$ (closed symbols) and $Cr_2Ge_2Te_6$/NiO (open symbols) with different thickness of $Cr_2Ge_2Te_6$ flakes. $Cr_2Ge_2Te_6$/NiO samples shows generally higher Curie temperatures than $Cr_2Ge_2Te_6$ over a wide range of the thickness of $Cr_2Ge_2Te_6$ for 20-nm-thick NiO (open circle) and 50-nm-thick NiO (open rectangle) as well as 35-nm-thick NiO (open diamond) and 100-nm-thick NiO (open triangle). Curie temperatures for $Cr_2Ge_2Te_6$ flakes without NiO are both from this study (closed circle) and from literature (diamond for [1] and triangle for [4]). Right most panel summarizes Curie temperatures of bulk $Cr_2Ge_2Te_6$ measured in this study by SQUID magnetometer and from literature (Carteaux et al [7], Zhang et al [8], Gong et al [1], Lin et al [9], Liu and Petrovic [10], Wang et al [4], Liu et al [11], Zeisner et al [12], Lohmann et al [13]). The results show that the Curie temperatures of $Cr_2Ge_2Te_6$/NiO with relatively thick (200 nm) $Cr_2Ge_2Te_6$ still shows higher Curie temperature (possibly contributed mostly by the $Cr_2Ge_2Te_6$/NiO interface) than bulk $Cr_2Ge_2Te_6$.